\documentstyle[12pt]{article}
\begin{document}
\font\bss=cmr12 scaled\magstep 0
\title{Closed dressing chains of D=1 and D=2 Toda lattice}
\author{A.V. Yurov\\
\small	Theoretical Physics Department,\\
\small	Kaliningrad State
University, \\
\small 236041, Al.Nevsky St., 14, Kaliningrad, Russia \\
\small	 email yurov@freemail.ru}
\date {}
\renewcommand{\abstractname}{\small Abstract}
\maketitle
\begin{abstract}
We apply the method of dressing chains to reproduction of Toda
lattice in the case of D=1 and D=2. On the example of modified
equations $m_0TL$ and $m_1TL$ it is shown that combination of the
Darboux and Schlesinger transformations results in closed dressing chains.
\end{abstract}
\thispagestyle{empty}
\medskip

\section {Introduction }

In this work we consider the method of dressing chains of
discrete symmetries for the Toda lattice (TL) (including the
two-dimensional generalization of them). One starts with nonlinear
Schrodinger equation (NLS)
$$
iu_t+u_{xx}+2u^2v=-iv_t+v_{xx}+2v^2u=0, \eqno(1)
$$
which admits two different types of discrete symmetries: Darboux
transformation (DT) and Schlesinger transformation (ST). The ST leads
to connection between NLS and TL models [1]. Using this circumstance and
Lax pair for the NLS model one can {\bf find} Lax pair for the TL one.
Then, using DT for the NLS and Lax pair for the TL one can obtain
the DT for the TL. Therefore total theory of  TL (Lax pair and
DT) can be constructed starting out from the theory of NLS.
Using  DT for the TL one can construct TL dressing chains  to
find modified TL equations which will be denoted as $m^1 TL$
(with superscript). To do that it is enough to know $L$-operator of
the NLS. By repeating this procedure with $A$-operator of the
equation (1) one find {\bf symmetries} of the $m^1 TL$ (the exact
definition of these symmetries is contained in [2]).  This plan is
realized in the Section 2. We'll show that it is appears the {\bf
closed} dressing chains which tie together TL and Volterra
equations, in contrast to open-ended dressing chains of the KdV
and NLS equations. This is the character  of the dressing chains
of the TL equations.

The alternative path to "multiply" the Toda lattice (see Sec. 3)
is as follows: One construct discrete symmetry chain for the NLS
to find modified NLS equations (we denote them by the $m_k NLS$).
Each of them inherits a ST from an initial NLS (1) which leads to
corresponding Toda-like lattice equations - $m_k TL$ (with
subscript). Since Lax pair and Darboux transformations for the
$m_k NLS$ are known (this is crucial point of the dressing chains
method) then one can find Lax pairs and DT for the $m_k TL$. Now,
if we apply the method from the previous paragraph then one will find
a new family of Toda-like equations (we denote them by the  as $m^n_kTL$.
The subscript points to the number of the modified NLS which
creates this $m^n_k TL$ equation. For example, all
Toda-like equations from the Sec. 2  belong to class  $m_0 TL$,
while all ones from the Sec. 3	belong to class  $m_1 TL$).

In Sec. 4 we generalize our approach to $D=2$ TL models using the
Davy-Stewartson equations (DS). Dressing chains of the first and second
types are constructed.

There are many works which are devoted to links between
integrable NLS-like models and difference-differential Toda-like
lattices (see, for example, [3]). The classification of
integrable lattice was made in [2], [4]. Our results are
associated with results in [5] and we are discussing that in Sec. 5.
This work is continuation of our research in [6], [7] which, for
one's turn, was inspired by the article [8].

\section{Equations  $m_0 TL$}

Lax pair for the (1) has the form
$$
\Psi_x=-i\sigma_3\Psi\Lambda+iS\Psi,\qquad
\Psi_t=-2i\sigma_3\Psi\Lambda^2+2iS\Psi\Lambda+W\Psi,\qquad
\eqno(2)
$$
where
$$
S=\left(\begin{array}{cc}
0&u\\
v&0
\end{array}\right),
\qquad \Lambda=\left(\begin{array}{cc}
\lambda&0\\
0&\mu
\end{array}\right),
\qquad W=\sigma_3\left(iS^2-S_x\right),
$$
$\lambda$ and $\mu$  are spectral parameters, $\sigma_3$  is the
third Pauli matrix, $\Psi$ -- $2\times 2$ matrix function. We
denote componentry of the first matrix $\Psi$ column by $\psi_n$,
 $\phi_n$, $u\equiv u_n$, $v\equiv v_n$. Then, the L-equation of
the system (2) has the form
$$
\psi_{n,x}=-i\lambda\psi_n+iu_n\phi_n,\qquad
\phi_{n,x}=i\lambda\phi_n+iv_n\psi_n. \eqno(3a-b)
$$
The components of the second $\Psi$ column are satisfying the
similar system with substitution  $\lambda\to\mu$.

The ST for the (1) are
$$
u_n\to u_{n+1}=u_n\left[u_nv_v+\left(\log
u_n\right)_{xx}\right],\qquad v_n\to v_{n+1}=\frac{1}{u_n},
\eqno(4a-b)
$$
$$
\psi_n\to\psi_{n+1}=\left(-2\lambda+i(\log
u_n)_x\right)\psi_n+u_n\phi_n,\qquad
\phi_n\to\phi_{n+1}=\frac{\psi_n}{u_n}, \eqno(5a-b)
$$
and reverse conversion:
$$
\begin{array}{l}
\displaystyle {v_n\to v_{n-1}=v_n\left[u_nv_v+\left(\log
v_n\right)_{xx}\right],\qquad
u_n\to u_{n-1}=\frac{1}{v_n},}\\
\displaystyle {\phi_n\to\phi_{n-1}=\left(2\lambda+i(\log
v_n)_x\right)\phi_n+v_n\psi_n,\qquad
\psi_n\to\psi_{n-1}=\frac{\phi_n}{v_n}.}
\end{array}
\eqno(6)
$$
Let us denote
$$
q_n\equiv\log(u_n),\qquad p_n\equiv q_{n,x},\qquad U_n\equiv
\frac{u_n}{u_{n-1}}=e^{q_n-q_{n-1}}. \eqno(7)
$$
It is easy to examine that functions $q_n$  are solutions of the
TL equations
$$
q_{n,xx}=e^{q_{n+1}-q_n}-e^{q_n-q_{n-1}}. \eqno(8)
$$
This remarkable point was noted in many works (see, for example,
[1]-[3]). Our first aim here is to construct Lax pair for the (8)
starting out from the equations (3). It is comfortably to
introduce the shift-operator $T$:
$$
Tu_n=u_{n+1}.
$$
We act $T^{-1}$ on the equation  (5b) then substitute the
expression for the $\phi_n$ into (3a) to obtain the first
equation of the Lax pair:
$$
\psi_{n,x}=-i\lambda\psi_n+iU_n\psi_{n-1}. \eqno(9)
$$
To obtain the second equation of the Lax pair we act $T$ onto the
(3b) and substitute into this expression $\phi_{n+1}$ (from the
(5b)), $v_{n+1}$ (from the (4b)) and $\psi_{n,x}$ from the (9).
As a result one get
$$
\psi_{n+1}=(2\lambda+ip_n)\psi_n+U_n\psi_{n-1}. \eqno(10)
$$
The consistency constraints of (9) and (10) has the form of two
equations
$$
p_{n,x}=U_{n+1}-U_n,\qquad U_{n,x}=U_n\left(p_n-p_{n-1}\right),
$$
which can be transformed into the  (8) by the (7). Thus, (9) and
(10) are nothing but Lax pair for the TL model.

Let us construct DT for the TL equation using {\bf elementary} DT
for the NLS  model (1) (see [9]). Let  $\psi_1$ and
$\phi_1$ be the components of the first matrix $\Psi$ column, where
$\Psi$ is the solution of the (2) with $\lambda=\lambda_1$. Then one
can write two sorts of the DT (the indices are omitted):
$$
\begin{array}{l}
\displaystyle
{\psi\to\psi^{(1)}=\left[2(\lambda-\lambda_1)+\frac{\phi_1}{\psi_1}u\right]\psi
-u\phi,\qquad \phi\to \phi^{(1)}=\phi-\frac{\phi_1}{\psi_1}\psi,}\\
\\
\displaystyle {u\to
u^{(1)}=iu_x-\left(2\lambda_1-\frac{\phi_1}{\psi_1}u\right)u,\qquad
v\to v^{(1)}=\frac{\phi_1}{\psi_1},}
\end{array}
\eqno(11)
$$
and
$$
\begin{array}{l}
\displaystyle
{\psi\to{}^{(1)}\psi=\psi-\frac{\psi_1}{\phi_1}\phi,\qquad
\phi\to{}^{(1)}\phi=\left[2(\lambda_1-\lambda)+\frac{\psi_1}{\phi_1}v\right]\phi
-v\psi,}\\
\\
\displaystyle {u\to {}^{(1)}u=\frac{\psi_1}{\phi_1},\qquad v\to
{}^{(1)}v=iv_x+\left(2\lambda_1+\frac{\psi_1}{\phi_1}v\right)v.}
\end{array}
\eqno(12)
$$
Let's $\psi_2$, $\phi_2$ be components of the second matrix $\Psi$
column with $\mu=\lambda_2$. Then one make
transformations  (11) and (12):
$$
u\to u^{(1)}\to {}^{(2)}u^{(1)},\qquad v\to v^{(1)}\to
{}^{(2)}v^{(1)},\qquad \psi\to \psi^{(1)}\to
{}^{(2)}\psi^{(1)},\qquad \phi\to \phi^{(1)}\to
{}^{(2)}\phi^{(1)}. \eqno(13)
$$
Note that the DT transformations are commutating
$$
{}^{(2)}u^{(1)}={}^{(1)}u^{(2)},\qquad
{}^{(2)}v^{(1)}={}^{(1)}v^{(2)}.
$$
If we'll write (13) explicitly for the twice dressed potentials and
wave functions it is easy to see that we would obtain the "ordinary"
DT for the NLS (see [10]). This is why we can call (11) and (12)
as elemantary DTs. Unfortunately these transformations are
uncomfortable in sense of getting exact solutions of the NLS because, in
the case of general position,  it is impossible to conserve
reduction restriction $u=\pm v^*$, in contrast to (13).

Now it is easy to find DT for the TL equations. Omitting simple
calculations we'll show the result. Let  $\{\psi_{1,n}\}$ be the
solution of the Lax pair (9), (10) with $\lambda=\lambda_1$. Then
we have two DTs for the TL equations which will be referred to as
$R$-transformation and $L$-transformation:
$$
R:\qquad\psi_n\to\psi_n^{(1)}=-\psi_{n+1}+\frac{\psi_{1,n+1}}{\psi_{1,n}}\psi_n,\qquad
q_n\to q_n^{(1)}=q_n+\log\frac{\psi_{1,n+1}}{\psi_{1,n}} \eqno(14)
$$
and
$$
L:\qquad\psi_n\to{}^{(1)}\psi_n=\psi_n-\frac{\psi_{1,n}}{\psi_{1,n-1}}\psi_{n-1},\qquad
q_n\to {}^{(1)}q_n=q_{n-1}+\log\frac{\psi_{1,n}}{\psi_{1,n-1}}.
\eqno(15)
$$

Using (9)-(10) it is quit easy to find first modified TL equations.
For this purpose one introduce new functions $\tau_n$ and $\xi_n$
$$
\tau_n=\frac{\psi_{1,n}}{\psi_{1,n+1}}=\frac{1}{\partial\xi_n},
$$
where $\partial\equiv -i\partial/\partial x$, $V_n\equiv ip_n$.
The shape of the function  $\tau_n$ is dictated from the  (14) and
(15). In new dependent variables, the Lax pair for the TL
equations has the form:
$$
\partial\tau_n=\tau_n\left(\tau_{n-1}U_n-\tau_{n}U_{n+1}\right),\qquad
\tau_{n+1}=\frac{1}{2\lambda_1+V_{n+1}+\tau_nU_{n+1}}. \eqno(16)
$$
Note that the first equations has quadratic nonlinearity by
fields $\tau_n$, as it should be in the method of the dressing
chains ([8]).

Eliminating potentials	$U_n$, $U_{n+1}$ and  $V_{n+1}$ from the
(16) one get equation $m^1TL$, which is just a Volterra
equation (one can write this in more customary	form; see (17a),
(52)):
$$
\partial^2\xi_n=\partial\xi_n\left(e^{\xi_n-\xi_{n+1}}-
e^{\xi_{n-1}-\xi_n}\right). \eqno(17)
$$

To obtain dressing chains of discrete symmetries one can use
either $R$ or $L$ transformations (14), (15). Let us choose the
(14). At the same time, the transformations regulation for the
$U_n$ and $V_n$ has the form:
$$
U_n\to U^{(1)}_n=\frac{\tau_{n-1}}{\tau_n}U_n,\qquad V_n\to
V^{(1)}_n=V_n+\tau_{n-1}U_n-\tau_nU_{n+1}.
$$
Let's take $\psi_{2,n}$  as the solution of  (9), (10)  with
$\lambda=\lambda_2$, $\psi_{2,n}^{(1)}$ is calculated from the
(14) whereas $\zeta_n$	is defined by the expression:
$$
\partial\zeta_n=\frac{\psi^{(1)}_{2,n+1}}{\psi^{(1)}_{2,n}}.
$$
After simple calculations one get these chains:
$$
e^{\zeta_n-\zeta_{n+1}}\,\partial\zeta_n=
e^{\xi_n-\xi_{n+1}}\,\partial\xi_{n+1},\qquad
\partial\left(\zeta_n-\xi_n\right)=
2(\lambda_2-\lambda_1)+e^{\zeta_{n-1}-\zeta_n}-e^{\xi_n-\xi_{n+1}}.
\eqno(18)
$$
(18) contain Lax pair for the (17). Denoting
$\zeta_n=\log\Psi_n$, we get

$$
\partial\Psi_n=A_n\Psi_{n+1}=\left(B_{n-1}+\tilde\lambda\right)\Psi_n+\Psi_{n-1},
\eqno(19)
$$
where
$$
A_n=e^{\xi_n-\xi_{n+1}}\,\partial\xi_{n+1},\qquad
B_n=\partial\xi_{n+1}-e^{\xi_{n+1}-\xi_{n+2}}, \eqno(20)
$$
and new spectral parameter $\tilde\lambda$ is connected with the
$\lambda$ by the proportion:
$\tilde\lambda=2(\lambda-\lambda_1)$. We note that the
consistency condition of  (19) has the form of usual Toda lattice
$$
\partial A_n=A_n(B_{n-1}-B_n),\qquad
\partial B_n=A_{n+1}-A_n
\eqno(21)
$$
and can be reduced to the  (17) by the substitution (20) into
the  (21). In new dependent variable the $R$-transformation (14)
is nothing else but  $L$-transformation (15). In other words, one
obtain the closed chains  of discrete symmetries which connect TL
equations with Volterra equations ($m^1 TL$).

To complete the picture we'll cite the corresponding formulas. Let
$\Psi_{1,n}$ be the solution of  (19) with
$\tilde\lambda=\tilde\lambda_1\ne 0$. $R$-transformation (14)
induce $L$-transformation (15) for the	(19) and (21) (one can
examine it by the direct calculation ):
$$
\begin{array}{l}
{\Psi_n\to{}^{(1)}\Psi_n=\Psi_n-\sigma_n\Psi_{n+1},\qquad
A_n\to {}^{(1)}A_n=A_n-\partial\sigma_n,}\\
\\
{B_n\to {}^{(1)}B_n=B_{n+1}+\partial\log\sigma_{n+1},}
\end{array}
$$
where  $\sigma_n=\Psi_{1,n}/\Psi_{1,n+1}$ are solutions of system
(compareto (16)):
$$
\partial\sigma_n=A_n-(B_n+\tilde\lambda_1)\sigma_n-\sigma_n^2,\qquad
\sigma_{n+1}=\frac{A_{n+1}}{B_n+\sigma_n+\tilde\lambda_1}.
\eqno(22)
$$
Using (21) and (22) we eliminate potentials $B_n$ and again
obtain Volterra equations:
$$
\partial\log\sigma_n=\beta_n-\beta_{n+1},\qquad
\partial\log\beta_n=\sigma_n-\sigma_{n-1},
\eqno(17a)
$$
where $\beta_n=A_n/\sigma_n$.

Analogously, one can consider t-equations. Rewriting (1) in new
variables $q_n$, $p_n$, and taking (8) into account we get the known
symmetry of the equations (8):
$$
-iq_{n,t}=p_n^2+e^{q_{n+1}-q_n}+e^{q_n-q_{n-1}}. \eqno(23)
$$
It is possible to repeat all actions  described above: to find
Lax pair, then to apply Darboux transformation to construct
symmetry of the Volterra equations. Resulting  formulas has been
described before (with the help of another approach; see [2],
[4]) so we omit them  here. Note that  [2]  contains the total
(up to gauge, linear, Galilean transformations) list of
integrable generalization of classical and relativist Toda
lattice in the form
$$
q_{n,xx}=\frac{1}{2}\left(G(q_{n+1},q_n,p_{n+1},p_n)-
F(q_n,q_{n-1},p_n,p_{n-1})\right),
$$
with symmetries $$
q_{n,t}=\frac{1}{2}\left(G(q_{n+1},q_n,p_{n+1},p_n)+
F(q_n,q_{n-1},p_n,p_{n-1})\right).
$$
Using the method of dressing chains we have found only two
equations from this list. In the next Section we'll obtain new
generalization of TL equations.
\section{Equations  $m_1 TL$}

Using DT (13) and Lax pair for the NLS (1) we can construct the
first modified NLS ($m_1 NLS$) which has the form
$$
\begin{array}{c}
\left(\beta^2-4bc\right)\left(ib_t+b_{xx}-2b^2c\right)+
2\alpha\left(\alpha c+2ic_x\right)b^2+
2(b_xc+2bc_x)b_x=0,\\
{}\\
\left(\beta^2-4bc\right)\left(-ic_t+c_{xx}-2c^2b\right)+
2\alpha\left(\alpha b-2ib_x\right)c^2+ 2(bc_x+2b_xc)c_x=0,
\end{array}
\eqno(24)
$$
where $\alpha=\lambda+\mu$, $\beta=\lambda-\mu$, $\lambda$ and
$\mu$ are spectral parameters of (2) as before, and $b=b(x,t)$
with $c=c(x,t)$ are new dependent variables which are defined by
$$
b=\frac{u-{}^{(2)}u^{(1)}}{2},\qquad
c=\frac{{}^{(2)}v^{(1)}-v}{2}.
$$
If  $\alpha=\beta=0$, then the equation (24) take  the elegant
form
$$
ib_t+b_{xx}-2b^2c-\frac{b_xc_x}{c}-\frac{1}{2}\frac{b_x^2}{b}=0,\qquad
-ic_t+c_{xx}-2bc^2-\frac{b_xc_x}{b}-\frac{1}{2}\frac{c_x^2}{c}=0.
\eqno(25)
$$
STs for the  (25) are defined by the formulas (compare to(4), (6)):
$$
\begin{array}{l}
\displaystyle {b_n\to b_{n+1}=\frac{1}{4}
\frac{(2c_nb_{n,x}^2-2c_nb_nb_{n,xx}+c_{n,x}b_{n,x}b_n+4c_n^2b_n^3)^2}
{c_n^2b_n(b_{n,x}^2-4c_nb_n^3)},\qquad
c_n\to c_{n+1}=\frac{4c_nb_n^2}{b_{n,x}^2-4c_nb_n^3},}\\
{}\\
\displaystyle {b_n\to
b_{n-1}=\frac{4b_nc_n^2}{c_{n,x}^2-4b_nc_n^3},\qquad c_n\to
c_{n-1}=\frac{1}{4}
\frac{(2b_nc_{n,x}^2-2b_nc_nc_{n,xx}+b_{n,x}c_{n,x}c_n+4b_n^2c_n^3)^2}
{b_n^2c_n(c_{n,x}^2-4b_nc_n^3)}},
\end{array}
\eqno(26)
$$
where
$$
(b_{n+1})_{n-1}=(b_{n-1})_{n+1}=b,\qquad
(c_{n+1})_{n-1}=(c_{n-1})_{n+1}=c.
$$

Formulas (26) were obtained from the discrete symmetries chains
of NLS in our work [7]. There is another (more spontaneous)
way to obtain them. This way is based on the comparative
analysis of (25) and (1).
Let's consider (for distinctness) the transformation  $n\to n+1$ in
(26). In the case of ST (4a-b) for the NLS one can see that new
fields are expressed through old ones via formulas (most of
indices are omitted in order to prevent jamming of formulas):
$$
u_{n+1}=U(u,v,u_x,u_{xx}),\qquad v_{n+1}=V(u). \eqno(27)
$$
By analogy with  (27), one need pick out for the (25)
$$
b_{n+1}=B(b,c,q,p,w),\qquad c_{n+1}=C(b,c,q), \eqno(28)
$$
where values $q=b_x$, $p=c_x$, $w=q_x$ must be considered as
independent variables whereas $B$ and $C$ should be obtained. Now
we should substitute (28) into mNLS then eliminate time
derivatives with the help of (25) and, at the end of this, to
equate to zero expessions attached to independent variables. In
such a way, equating to zero the factor attached to  $w_x$ in the
second equation (25) we get (after simple integration by $w$):
$$
B=\left(wC_q+qC_b+pC_c\right)^2S(b,c,q,p),
$$
where $S(b,c,q,p)$ is arbitrary function of it's  arguments.
By equating to zero the factor attached to $w$ one get
$$
C_q=0,\qquad S_p=0.
$$
The further inquiry  show that we need to choose the second way
so $S=Z(b,c,q)$. Then, the nulling of the factor attached to
$p_x$ results in simple PDE and the  integration of this equation
gives us the functional dependence for the $C$:
$$
C=C\left(b,\frac{q^2}{c}\right).
$$
The factor attached to $w^2$ leads to  Riccati equation (that's a bad news)
but we can simplify the problem (here are a good ones) using  the  following
observation: Let
$$
b=g_1(t)+\epsilon f_1(x,t),\qquad c=g_2(t)+\epsilon f_2(x,t).
\eqno(29)
$$
Substituting (29) into the (25) we see that if	$\epsilon\to 0$
then two last items in (25) are negligible quantities so the
equations (25) are reduced to (non-soliton) NLS. This implies
that if $\epsilon\to 0$ then our desired ST must be reduced into
ST (4). Therefore
$$
\frac{1}{C}=-b+F\left(b,\frac{q^2}{c}\right),
$$
where $F(b,0)=0$ ($\epsilon\to 0$ mean that  $q\to 0$ and $p\to
0$). The last condition means that for the $F$ one can use Taylor
with respect to $q^2/c$. Since the dimensionalities of	$b$, $c$
and $x$ are connected by the proportion $[b][c]=[x]^{-2}$, we get
the initial object for the $C$:
$$
\frac{1}{C}=-b+\sum_{m=1}^{\infty}G_m\left(\frac{q^2}{c}\right)^m
b^{1-3m}, \eqno(30)
$$
where $G_m$  are dimensionless constants. In fact , it is enough
to restrict ourself to the first term of series in (30) (we can
verify, comparing (30) with (26), that	$G_1=1/4$ whereas the
rest coefficients are zero). Using this obtained initial object
one can continue  the calculation to find (26).

If  $\alpha$ and $\beta$ are nonvanishing numbers then
$$
\begin{array}{l}
\displaystyle {b_{n-1}=
\frac{ic_n(\beta^2-4b_nc_n)}{i(\alpha^2-\beta^2)c_n^2-2\alpha c_nc_{n,x}+i(4c_n^3b_n-c_{n,x}^2)},}\\
\\
\displaystyle {c_{n-1}=\frac{Z(-\alpha,\beta,c_n,b_n)}
{\left(\beta^2-4bc\right)^2\left[i(\alpha^2-\beta^2)c_n^2-2\alpha
c_nc_{n,x}+ i(4c_n^3b_n-c_{n,x}^2)\right]},}
\end{array}
\eqno(31)
$$
$$
\begin{array}{l}
\displaystyle {b_{n+1}=\frac{Z(\alpha,\beta,b_n,c_n)}
{\left(\beta^2-4bc\right)^2\left[i(\alpha^2-\beta^2)b_n^2+
2\alpha b_nb_{n,x}+i(4b_n^3c_n-b_{n,x}^2)\right]}},\\
\\
\displaystyle {c_{n+1}=
\frac{ib_n(\beta^2-4b_nc_n)}{i(\alpha^2-\beta^2)b_n^2+ 2\alpha
b_nb_{n,x}+i(4b_n^3c_n-b_{n,x}^2)}.}
\end{array}
\eqno(32)
$$
The quantity  $Z(\alpha,\beta,b,c)$ has bulky form so we'll show
it in two extreme cases when either $\beta=0$ or $\alpha=0$:
$$
\begin{array}{l}
\displaystyle
{\frac{Z(\alpha,0,b,c)}{4b}=}\\
\displaystyle {-i(b^2c\alpha^2)^2+2b^3c(bc_x-b_xc)\alpha^3-
ib^2\left[3(b_xc)^2+(3bc)^3+4bc(b_xc_x-cb_{xx})-(bc_x)^2\right]\alpha^2+}\\
\displaystyle
{2b(bc_x-b_xc)\left((2cb)^2b+b_xc_xb-2bcb_{xx}+2b_x^2c\right)\alpha-
i\left((2cb)^2b+b_xc_xb-2bcb_{xx}+2b_x^2c\right)^2,}\\
\\
\displaystyle
{-iZ(0,\beta,b,c)=}\\
\displaystyle
{b(b^3c+b_x^2-bb_{xx})\beta^6+\left[(b_x^2-bb_{xx})b_{xx}-2b^3b_xc_x-
12b^2c\left(b^3c+b_x^2-bb_{xx}\right)\right]\beta^4+}\\
\displaystyle
{\left[48b^3c^2\left(b_x^2+b^3c-bb_{xx}\right)+8c(bb_{xx})^2-4bb_x(3b_xc+bc_x)b_{xx}+
(3cb_x^2+2bc_xb_x^2+16b^4cc_x)b_x\right]\beta^2-}\\
\displaystyle
{-4b\left((2cb)^2b+b_xc_xb-2bcb_{xx}+2b_x^2c\right)^2}.
\end{array}
$$
Formulas (31), (32) result in new Toda-like lattice. Denoting
$$
P_n=\log b_n,\qquad Q_n=\log c_n,
$$
one find the equations	$m_1^0TL\equiv m_1TL$ in familiar
Euclidean variables:
$$
\begin{array}{l}
\displaystyle {(\partial P_n)^2+2\alpha\partial P_n+4e^{ P_n+Q_n}-
\beta^2e^{-P_n-Q_{n+1}}+
4e^{Q_n-Q_{n+1}}=\beta^2-\alpha^2,}\\
\\
\displaystyle {(\partial Q_n)^2-2\alpha\partial Q_n+4e^{ P_n+Q_n}-
\beta^2e^{-Q_n-P_{n-1}}+ 4e^{P_n-P_{n-1}}=\beta^2-\alpha^2.}
\end{array}
\eqno(33)
$$
Thus we have integrable lattice composed of interaction nodes
(atoms) of two kinds. It is  most simple case when
$\alpha=\beta=0$. In this case each equation looks as a law of each atom's
total energy conservation.  In other words, in this
case the equations (33) describes zero-point oscillations of
lattice.

Lax pair for the  (33) can be obtained via NLS discrete
symmetries chains. As a result one get
$$
\begin{array}{l}
\displaystyle
{(\alpha-2a_n)\partial\Psi_n+2(\alpha_1-A_n)(\alpha-2a_n)\Psi_n-
(\alpha_1-2A_n)(\partial b_n+2a_nb_n)=0,}\\
\\
\displaystyle
{\Psi_{n-1}=\frac{\left(\beta_1^2-4\Psi_n\Phi_n\right)\Phi_n}
{\left(\alpha_1^2-\beta_1^2\right)\Phi_n^2-2\alpha_1\Phi_n\partial\Phi_n+
4\Phi_n^3\Psi_n+\left(\partial\Phi_n\right)^2},}\\
\\
\displaystyle
{(\alpha-2a_n)\partial\Phi_n+2\left((\alpha-a_n)\alpha_1+(4a_n-3\alpha)A_n\right)
\Phi_n-(\alpha-2A_n)\partial c_n=0,}\\
\\
\displaystyle
{\Phi_{n+1}=\frac{\left(\beta_1^2-4\Psi_n\Phi_n\right)\Psi_n}
{\left(\alpha_1^2-\beta_1^2\right)\Psi_n^2+2\alpha_1\Psi_n\partial\Psi_n+
4\Psi_n^3\Phi_n+\left(\partial\Psi_n\right)^2},}
\end{array}
\eqno(34)
$$
where $\Psi_n$, $\Phi_n$ are wave functions of the spectral
problem, $\alpha_1$, $\beta_1$	are spectral parameters
($\alpha$ and  $\beta$ are fixed) whereas  functions $a_n$,
$A_n$ are defined as:
$$
a_n=\frac{1}{2}\left(\alpha\pm\sqrt{\beta^2-4b_nc_n}\right),\qquad
A_n=\frac{1}{2}\left(\alpha_1\pm\sqrt{\beta_1^2-4\Psi_n\Phi_n}\right).
$$
Equations (33), (34) are analogue of Lax pair (16) for the TL equations
(8). It is quite obvious from (34) that  $\log\Psi_n$, $\log\Phi_n$ are
solutions of the same $m_1 TL$ equations (33) with the change
$\alpha\to \alpha_1$, $\beta\to\beta_1$.

At last, using (13) it is easy to calculate corresponding Darboux
transformation for the (34). One can predict the response:
$$
b_n\to \Psi_n,\qquad c_n\to \Phi_n.
$$
Intermediate equations $m_1^2 TL$ (i.e. the analogue of the Volterra equation for the (33)) can be obtained from the above mentioned formulas:
$$
\hat{\Psi}_n=\Psi_n-b_n,\qquad \hat{\Phi}_n=\Phi_n-c_n. \eqno(35)
$$
The Lax pair for the $m_1^2 TL$  equation
can be obtained from the (34) by
the elimination of   $\Psi_n$ and  $\Phi_n$ with the help of the
(35). In the role of wave function one get $b_n$ and  $c_n$
(with another values of $\alpha$,  $\beta$). Thus, we get
closed dressing chains again.
\section{D=2 TL equations}

Above-stated formalism is also work  for the $D=2$ TL equations. These
equations can be obtained from the DS equations:
$$
\begin{array}{l}
\displaystyle
{iu_t+u_{xx}+\frac{1}{\alpha^2}u_{yy}-\frac{2}{\alpha^2}u^2v+gu=0,\qquad
-iv_t+v_{xx}+\frac{1}{\alpha^2}v_{yy}-\frac{2}{\alpha^2}v^2u+gv=0},\\
{}\\
g_{yy}-\alpha^2g_{xx}=-4\left(uv\right)_{xx}.
\end{array}
\eqno(36)
$$
Here  $\alpha^2=\pm 1$. The Lax "pair"  for the  (36) is the
system of four scalar equations,
$$
\begin{array}{l}
\psi_y=\alpha\psi_x+u\phi,\qquad
\phi_y=-\alpha\phi_x+v\psi,\\
{}\\
\displaystyle {\psi_t=2i\psi_{xx}+\frac{2i}{\alpha}u\phi_x+
\left(\frac{1}{2}\left[\frac{1}{\alpha}F_y+F_x\right]-\frac{i}{\alpha^2}uv
\right)\psi+\frac{i}{\alpha^2}\left(\alpha u_x+u_y\right)\phi,}\\
{}\\
\displaystyle {\phi_t=-2i\phi_{xx}+\frac{2i}{\alpha}v\psi_x+
\left(\frac{1}{2}\left[\frac{1}{\alpha}F_y-F_x\right]+\frac{i}{\alpha^2}uv
\right)\phi+\frac{i}{\alpha^2}\left(\alpha v_x-v_y\right)\psi,}
\end{array}
\eqno(37)
$$
where $g=-iF_x$.

Now let two twains of functions $\{\psi_1,\,\phi_1;\,\psi,\,\phi\}$
be solution of the (37) with any given $u$, $v$ and $F$. Elementary
Darboux transformations for the DS equations (i.e. the analogue
of the (11), (12))  has the form,
$$
\begin{array}{l}
\displaystyle
{\psi\to\psi^{(1)}=-2\alpha\psi_x-u\phi+\left(2\alpha\psi_{1,x}+u\phi_1\right)
\frac{\psi}{\psi_1},
\qquad \phi\to \phi^{(1)}=\phi-\frac{\phi_1}{\psi_1}\psi,}\\
\\
\displaystyle {u\to
u^{(1)}=\frac{u}{\psi_1}\left(2\alpha\psi_{1,x}+u\phi_1\right)-u_y-\alpha
u_x, \qquad
v\to v^{(1)}=\frac{\phi_1}{\psi_1},}\\
\\
\displaystyle {F\to F^{(1)}=F+4i\frac{\psi_{1,x}}{\psi_1}}
\end{array}
\eqno(38)
$$
and
$$
\begin{array}{l}
\displaystyle {\phi\to
{}^{(1)}\phi=2\alpha\phi_x-v\psi-\left(2\alpha\phi_{1,x}-v\psi_1\right)
\frac{\phi}{\phi_1},
\qquad \psi\to{}^{(1)}\psi=\psi-\frac{\psi_1}{\phi_1}\phi,}\\
\\
\displaystyle {v\to
{}^{(1)}v=\frac{v}{\phi_1}\left(v\psi_1-2\alpha\phi_{1,x}\right)-v_y+\alpha
v_x, \qquad
u\to {}^{(1)}u=\frac{\psi_1}{\phi_1},}\\
\\
\displaystyle {F\to {}^{(1)}F=F+4i\frac{\phi_{1,x}}{\phi_1}}.
\end{array}
\eqno(39)
$$
Two-dimensional problem has larger list of discrete symmetries
then single-dimensional one. The new type of them is so called binary DT which can
be obtained as follows: DS equations has  Lax pair
which is "conjugate" to (37),
$$
\begin{array}{l}
p_y=\alpha p_x-vf,\qquad
f_y=-\alpha f_x-up,\\
{}\\
\displaystyle {p_t=-2ip_{xx}+\frac{2i}{\alpha}vf_x-
\left(\frac{1}{2}\left[\frac{1}{\alpha}F_y+F_x\right]-\frac{i}{\alpha^2}uv
\right)p+\frac{i}{\alpha^2}\left(\alpha v_x+v_y\right)f,}\\
{}\\
\displaystyle {f_t=2if_{xx}+\frac{2i}{\alpha}up_x-
\left(\frac{1}{2}\left[\frac{1}{\alpha}F_y-F_x\right]+\frac{i}{\alpha^2}uv
\right)f+\frac{i}{\alpha^2}\left(\alpha u_x-u_y\right)p.}
\end{array}
\eqno(40)
$$
For  accommodation we denote wave function of Lax pair (40) using
Roman alphabet $p$, $f$ (Greek letters $\psi$, $\phi$ will be
used for the Lax pair  (37)), and DT for the (40) will be marked
with the help of subscripts in round paranthesis:
$$
\begin{array}{l}
\displaystyle {p\to p_{_{(1)}}=2\alpha p_x-vf-\left(2\alpha
p_{1,x}-vf_1\right) \frac{p}{p_1},
\qquad f\to f_{_{(1)}}=f-\frac{f_1}{p_1}p,}\\
\\
\displaystyle {v\to v_{_{(1)}}=\frac{v}{p_1}\left(vf_1-2\alpha
p_{1,x}\right)+v_y+\alpha v_x, \qquad
u\to u_{_{(1)}}=\frac{f_1}{p_1},}\\
\\
\displaystyle {F\to F_{_{(1)}}=F+4i\frac{p_{1,x}}{p_1}},
\end{array}
\eqno(41)
$$
$$
\begin{array}{l}
\displaystyle {f\to {}_{_{(1)}}f=-2\alpha f_x-up+\left(2\alpha
f_{1,x}+up_1\right) \frac{f}{f_1},
\qquad p\to{}_{_{(1)}}p=p-\frac{p_1}{f_1}f,}\\
\\
\displaystyle {u\to {}_{_{(1)}}u=\frac{u}{f_1}\left(up_1+2\alpha
f_{1,x}\right)+u_y-\alpha u_x, \qquad
v\to {}_{_{(1)}}v=\frac{p_1}{f_1},}\\
\\
\displaystyle {F\to {}_{_{(1)}}F=F+4i\frac{f_{1,x}}{f_1}}.
\end{array}
\eqno(42)
$$
DT  (38)-(39) result in following transformations laws for the
solutions of the system (40):
$$
\begin{array}{l}
\displaystyle {p\to
p^{(1)}=\frac{A+\Omega(\psi_1,\phi_1;p,f)}{\psi_1},\qquad
f\to f^{(1)}=2\alpha f+\frac{u}{\psi_1}\left(A+\Omega(\psi_1,\phi_1;p,f)\right),}\\
\\
\displaystyle {f\to
{}^{(1)}f=\frac{A+\Omega(\psi_1,\phi_1;p,f)}{\phi_1},\qquad p\to
{}^{(1)}p=-2\alpha
p+\frac{v}{\phi_1}\left(A+\Omega(\psi_1,\phi_1;p,f)\right),}
\end{array}
\eqno(43)
$$
where
$$
\begin{array}{l}
\displaystyle
{\Omega(\psi,\phi;p,f)=\int\,d\Omega(\psi,\phi;p,f),\qquad
d\Omega(\psi,\phi;p,f)=(\psi p+\phi f)dx+\alpha(\psi p-\phi f)dy+}\\
\\
\displaystyle {+2i\left[\frac{1}{\alpha}\left(v\psi f+u\phi
p\right)+p\psi_x-p_x\psi+ f_x\phi-f\phi_x\right]dt,}
\end{array}
$$
$A$  is arbitrary constant, 1-form $d\Omega(\psi,\phi;p,f)$ is
closed if $\psi$, $\phi$, $p$ and $f$ are solutions of the  (37),
(40) with the same potentials $u$, $v$ and $F$. Similar formulas
for  $\psi_{_{(1)}}$, $\phi_{_{(1)}}$, ${}_{_{(1)}}\psi$,
${}_{_{(1)}}\phi$ which are induced by the transformations
(41)-(42) have the form:
$$
\begin{array}{l}
\displaystyle {\psi\to
\psi_{_{(1)}}=\frac{A+\Omega(\psi,\phi;p_1,f_1)}{p_1},\qquad
\phi\to \phi_{_{(1)}}=-2\alpha\phi+\frac{v}{p_1}\left(A+\Omega(\psi,\phi;p_1,f_1)\right),}\\
\\
\displaystyle {\phi\to
{}_{_{(1)}}\phi=\frac{A+\Omega(\psi,\phi;p_1,f_1)}{f_1},\qquad
\psi\to
{}_{_{(1)}}\psi=2\alpha\psi+\frac{u}{f_1}\left(A+\Omega(\psi,\phi;p_1,f_1)\right).}
\end{array}
\eqno(44)
$$
Using  (38), (39), (41-44) one can find elemantary binary DT.
These transformations allow us to calculate potentials with two
indices: one on top and one below. For example
$$
u_{_{(1)}}^{_{(1)}}={}_{_{(1)}}^{_{(1)}}u=\frac{1}{{}_{_{(1)}}v^{_{(1)}}}=
u+\frac{2\alpha f_1\psi_1}{A+\Omega_1},\qquad
v_{_{(1)}}^{_{(1)}}={}_{_{(1)}}^{_{(1)}}v=\frac{1}{{}^{_{(1)}}u_{_{(1)}}}=
v-\frac{2\alpha p_1\phi_1}{A+\Omega_1}, \eqno(45)
$$
where  $\Omega_1=\Omega(\psi_1,\phi_1;p_1,f_1)$. We'll not write
out  these formulas.
\newline
\newline
{\em Remark.} The twain  of linear equations for the $\psi_y$,
$\phi_y$ (37) (or for the $p_y$, $f_y$ (40)) can be represented
as the single linear second-order equation with variable
coefficients. Imposing the special restrictions on the potentials $u$,
$v$, one can obtain  some famous  equations, e.g.
Laplace-Moutard  equation [11] or the Goursat one [12]. To obtain
the  Goursat equation one need to choose
$$
v=\pm u. \eqno(46)
$$
The usual Darboux transformations don't conserve the reduction
restrictions
 (46), while the elementary binary transformations (45) conserve them by the choice
$$
p=\psi,\qquad f=\mp\phi,
$$
which don't conflict with (37), (40). As a result we get a well
known analog of a Moutard transformation for the Gourst equation
(see, for example, [13], [14), which is nothing but outcome of two
successive elementary DT. Similarly it is possible to obtain
usual Moutard transformation. Note that this transformation is
the useful method to construct explicit solutions of the
Veselov-Novikov equation  [10], [15]. In turn the Goursat equation
produce $D=2$ mKdV hierarchy so transformations (45) (which don't
conflict with (46)) can be used to construct exact solutions of
equations from this hierarchy.

The DS equations (36) admits ST ([7], [16]):
$$
\begin{array}{l}
\displaystyle {u_n\to u_{n+1}=u_n\left(u_nv_n+\alpha^2(\log
u_n)_{xx}-(\log u_n)_{yy}\right),\qquad
v_n\to v_{n+1}=\frac{1}{u_n},}\\
g_n\to g_{n+1}=g_n+4(\log u_n)_{xx}\\
{}\\
\displaystyle {u_n\to u_{n-1}=\frac{1}{v_n},\qquad
v_n\to v_{n-1}=v_n\left(u_nv_n+\alpha^2(\log v_n)_{xx}-(\log v_n)_{yy}\right),}\\
g_n\to g_{n-1}=g_n+4(\log v_n)_{xx},
\end{array}
\eqno(47)
$$
from which one get the equations of $D=2$ Toda lattice:
$$
\alpha^2q_{n,xx}-q_{n,yy}=e^{q_{n+1}-q_n}-e^{q_n-q_{n-1}},\qquad
q_n=\log(u_n). \eqno(48)
$$
Using (48) and (36) we find symmetries of the (48) (see. (23):
$$
-iq_{n,t}=2q_{n,xx}+q_{n,x}^2+\frac{1}{\alpha^2}q_{n,y}^2-
\frac{1}{\alpha^2}\left(e^{q_{n+1}-q_n}+e^{q_n-q_{n-1}}\right)+g_n,
$$
where $g_n$ can be expressed as:
$$
g_n(x,y,t)=\tilde g_n(x,y)+
4\int_{-\infty}^{\infty}dx'dy'G(x',y';x,y)\left(e^{q_n-q_{n-1}}\right)_{x'x'},
$$
where $G(x',y';x,y)$  is the Green function
satisfying  the equation
$$
\left(\partial_y^2-\alpha^2\partial_x^2\right)G(x',y';x,y)=
-\delta(x-x')\delta(y-y'),
$$
where $\tilde{g_n}$ is the solution of this equation with a zero
second member of this one.

The plan of searching of Lax pair, Darboux transformations and
dressing chains for the (48) is the same as  the plan in Sec. 2:
Using  (37), (40)  and	(47) one find two "conjugate" Lax pairs
for the (48):
 $$
\begin{array}{l}
\displaystyle
{\psi_{n,y}=\alpha\psi_{n,x}+e^{q_n-q_{n-1}}\psi_{n-1},}\\
\\
\displaystyle {\psi_{n+1}=2\alpha\psi_{n,x}-(\alpha
q_{n,x}+q_{n,y})\psi_n+ e^{q_n-q_{n-1}}\psi_{n-1},}
\end{array}
\eqno(49)
$$
and
$$
\begin{array}{l}
\displaystyle
{f_{n,y}=-\alpha f_{n,x}-e^{q_n-q_{n-1}}f_{n-1},}\\
\\
\displaystyle {f_{n+1}=2\alpha f_{n,x}+(q_{n,y}-\alpha
q_{n,x})f_n+ e^{q_n-q_{n-1}}f_{n-1}.}
\end{array}
\eqno(50)
$$
Using now DT (38-39) and (41-42) it is easy to obtain requisite
formulas for the (49), (50).  The result is just the same as
$R$- and  $L$-transformations (14-15), and this is true for both the
(49) and (50). The last statement  is obvious  because the Lax
pair (50) can be obtained from the (49) by the substitute
$$
\alpha\to-\alpha,\qquad \psi_{n+k}\to f_{n+k}=(-1)^k\psi_{n+k}.
$$
Starting out from the (49) one get $D=2$ Volterra equations
(compare to (17)):
$$
\xi_{n,_{XY}}=\xi_{n,_{X}}\left(e^{\xi_n-\xi_{n+1}}-
e^{\xi_{n-1}-\xi_n}\right), \eqno(51)
$$
where
$$
\partial_{_X}=\partial_y+\alpha\partial_x,\qquad
\partial_{_Y}=\partial_y-\alpha\partial_x,\qquad
\xi_{n,_{X}}=\frac{\psi_{1,n+1}}{\psi_{1,n}},
$$
where $\psi_{1,n}$  is some particular solution of the	(49).
Introducing  new dependent variable $a_n$ and $b_n$:
$$
a_n=\xi_{n,_{X}},\qquad b_n=e^{\xi_n-\xi_{n+1}},
$$
it is possible to present (51) in more customary form
$$
\left(\log a_n\right)_{_Y}=b_n-b_{n-1},\qquad \left(\log
b_n\right)_{_X}=a_n-a_{n+1}. \eqno(52)
$$
And If we start from the (50) then the resulting equations will be
gage-equivalent to (51). But we'll not apply them here.

Finally, let apply  dressing chains. Choosing Lax pair (49) and
$R$-transformation  (14) (it can be obtained from the (38)) we
have:
$$
e^{\zeta_n-\zeta_{n+1}}\,\zeta_{n,_{X}}=
e^{\xi_n-\xi_{n+1}}\,\xi_{n+1,_{X}},\qquad
\left(\zeta_n-\xi_n\right)_{_Y}=
e^{\zeta_{n-1}-\zeta_n}-e^{\xi_n-\xi_{n+1}}, \eqno(53)
$$
where  $\zeta_n$ is defined just as in Sec. 2.

Chains (53) are $D=2$ generalization of chains (18). In addition
, using binary DT we can construct dressing chains of the second
type. Denoting (see (43))
$$
\eta_{n,_{Y}}=\frac{f^{(1)}_{n+1}}{f^{(1)}_n},
$$
one get
$$
e^{\eta_{n+1}-\eta_n}\,\eta_{n,_{Y}}=
e^{\xi_{n}-\xi_{n+1}}\,\xi_{n+1,_{X}},\qquad
\left(\eta_{n,_{X}}-e^{\eta_n-\eta_{n-1}}\right)_{_Y}=
\left(e^{\xi_n-\xi_{n+1}}-\xi_{n,_{Y}}\right)_{_X}. \eqno(54)
$$
Eliminating  $\xi_n$ from the  (54) we obtain nonlinear equation
for the functions $\eta_n$, whereas chains (54) can be considered
as Lax pair for this equation.

\section{Conclusion}

Main results of this work are:
\newline
\newline
1. Dressing chains for the Toda lattice are constructed starting
out from the NLS theory.
\newline
2. It is shown that dressing chains for the Toda lattice are
closed.
 \newline
3.The methods allowing to construct the $m^n_k TL$ equations are
suggested. The particular example of equation from this set
(including it's Lax pair, (33-34)) is studied.
\newline
4. Dressing chains of the first and second type for the $D=2$
Toda equations are obtained.

The links between NLS and TL were studied in [5]. Instead of
traditional system of Zakharov-Shabat (3),  authors of [5] have
used the  single second-order equation
$$
\Psi_{xx}+(z-2i\lambda)\Psi_x+p\Psi=0, \eqno(55)
$$
which can be obtained from the (3) if  one put
$$
z=-(\log u)_x,\qquad p=uv,\qquad \Psi=e^{i\lambda x}\psi.
$$
Symmetries from the [5] are nothing but ST (6) (in [5] these
transformations were called T-transformations) and DT
(S-transformations in [5]). According to lemma which was proved
in [5],  S- and T-transformations are unique iso-spectral
symmetries in the form	$\Psi\to f\Psi_x+g\Psi$,  where
coefficients  $f$ and $g$  are independent from the spectral
parameter $\lambda$. It is obviously that transformations (4-5)
and (11) don't satisfy this demand. One can show that they  are
connected with transformations $T^{-1}$ and $S^{-1}$. It is
obvious in the case (4-5). The link between $S^{-1}$ and (11) has
more delicate nature, so let consider it in detail.

The straightforward calculation by the (12) result in
S-transformation for the equation (55):
$$
\Psi\to {}^{(1)}\Psi=\Psi-\frac{\Psi_x}{\Psi_{1,x}}\Psi_1,
\eqno(56)
$$
where $\Psi_1$	is the solution of the	(55) with
$\lambda=\lambda_1$.

Let us find the inverse transformation,  acting accordingly  to
the  [5]: Differentiating (56) and excluding  $\Psi_x$ we get,
$$
\Psi=\Psi\left({}^{(1)}\Psi,{}^{(1)}\Psi_x;\Psi_1,\Psi_{1,x}\right).
\eqno(57)
$$
but it is not the ultimate result. That is because during the
inversion of the formula (56) one must have in mind that second
member (57) must  be expressed via "dressed" values  only (i.e.
via values with left superscript "(1)").  To lead (57) to this
form we need to introduce solution $\hat\Psi$ which will be
linear-independent with $\Psi$.  The general solution of (55)
(with fixed $\lambda$) will be linear combination of  $\Psi$ and
$\hat\Psi$ with any given constant coefficients.

It is easy to find the function $\hat\Psi$ by the standard
algorithm (to multiply (55) by the $\hat\Psi$; to multiply the equation for the
$\hat\Psi$ by the $\Psi$; to subtract and to integrate).
Then one can dress   $\hat\Psi_1$  (for the  $\lambda=\lambda_1$)
by the formula	(56) (we need to use this roundabout way to
escape the result ${}^{(1)}\Psi_1=0$, which we have faced by the
direct application of the (57)):
$$
\hat\Psi_1\to{}^{(1)}\hat\Psi_1.
$$
We differentiate this formula, then we express $\Psi_1$ and
$\Psi_{1,x}$ via ${}^{(1)}\hat\Psi_1$, ${}^{(1)}\hat\Psi_{1,x}$
and substitute it into the (57). Now it is possible to determine
that (57) is gage- equivalent  to (11). Let  $\psi$ and $\phi$
be solutions of the  (3). The necessary  for us
linear-independent functions are:
$$
\hat\psi(x)=\frac{1}{2}\psi(x)\int dz\,
sgn(x-z)\,\frac{u(z)}{\psi^2(z)},\qquad
\hat\phi(x)=\frac{1}{2}\phi(x)\int dz\,
sgn(x-z)\,\frac{u(z)}{\psi^2(z)}- \frac{i}{\psi}.
$$
Dressing these expessions by the formulas (12) and substituting
$\lambda=\lambda_1$, $\psi=\psi_1$, $\phi=\phi_1$, one get
$$
{}^{(1)}\hat\psi_1=\frac{i}{\phi_1},\qquad
{}^{(1)}\hat\phi_1=-\frac{iv}{\phi_1}.
$$
Finally, substituting obtained functions into the (11) we have
$$
{}^{(1)}u\to -u,\qquad {}^{(1)}v\to -v,
$$
Q.E.D.

The calibration (55) from the	[5] is suitable  to research
closure conditions  and for construction of  potentials  which are
invariant with respect to S- and T-transformations. In contrast
to [5], our purpose here is the constructions of the dressing
chains, in the manner of [8]. I conclude that the standard
symmetrical calibration of Zakharov-Shabat is more comfortable
and natural for this goal.

%\vfill
%\eject

%\newpage
$$
{}
$$
{\em Acknowledgements.}

The author are grateful to V. Yurov
for useful comments and his help.
This work was partially supported by the Grant of Education Department of the Russian Federation,
 No. E00-3.1-383.
$$
{}
$$
\centerline{\bf References} \noindent
\begin{enumerate}
\item  S.I. Svinolupov and  R.I. Yamilov. \rm\, Theor. Math.Phys. (1994) V.98. N2. P.207.
\item  V.A. Adler and  A.B. Shabat.\rm\, Theor. Math.Phys. (1997) V.111. N3. P.323.
\item  D. Levi. \rm\, J. Phys. A: Math. Gen. (1981) V.14. P.1083; A.B.
Shabat and  R.I. Yamilov.\rm\, Algebra and analysis. (1990) V.2. 2. P.183;
A.N. Leznov,A.B. Shabat, R.I. Yamilov. \rm \,
Phys.Lett.A. (1993) V.174. P.397.
\item  R.I. Yamilov. \rm\, Uspeh. Math. Nauk. (1983) V.38. V.6. P.155;
V.A. Adler and	A.B. Shabat.\rm\, Theor. Math.Phys. (1997) V.112. N2. P.179;
V.A. Adler and	A.B. Shabat.\rm\, Theor. Math.Phys. (1998) V.115. N3. P.349.
\item  A.B. Shabat.\rm\, Theor. Math.Phys. (1999) V.121. N1. P.165.
\item	A.V. Yurov. \rm\,  Theor. Math.Phys. (1999) V.119. N3. P.419.
\item  A.V. Yurov.  \rm\, J. Math. Phys. (2003) V.44. N3. P.1183
[nlin.SI/0207001].
\item  B.A. Borisov and  S.A. Zykov.\rm\,  Theor. Math.Phys. (1998)
V.115. N2. P.199.
\item  S.B. Leble, N.V. Ustinov.\rm\, J. Math. Phys. (1993) V.34. P.1421.
\item	V.B. Matveev, M.A. Salle.\rm\,	Darboux Transformation and
    Solitons. Berlin--Heidelberg: Springer Verlag, 1991.
\item  T. Moutard. \rm\, J. Ecole Polytechnique. (1878) 45. P.1.
\item  E. Goursat. \rm\, Bull. Soc. Math. France. (1897) V.25. P.36.
\item  E.I. Ganzha. \rm\, $On\,\, One\,\, Analogue\,\, of\,\, the\,\,
Moutard\,\, Transformation\,\, for\,\, the\, Goursat$\,\,
$Equation\,\, \theta_{xy}=2\sqrt{\lambda(x,y)\theta_x\theta_y}.$
solv-int @ xyz.lant.gov No 9806001.
\item S.B. Leble and A.V. Yurov \rm\, J. Math. Phys. (2002) V.43. N2. P.1095.
\item  C. Athorne, J.J. Nimmo. \rm\, Inverse Problems. (1991) 7. P.809.
\item	A.V. Yurov. \rm\,  Theor. Math.Phys. (1996) V.109. N3. P.338.

\end{enumerate}

\vfill
\eject

\end{document}